\documentclass[aps,prd,twocolumn,groupedaddress,showpacs,nofootinbib,amssymb,balancelastpage,preprintnumbers,floatfix]{revtex4}

\usepackage[dvipdfmx]{graphicx}
\usepackage{graphicx,bm,color}

\usepackage{amsmath}
\usepackage{amssymb}
\usepackage{amsfonts}
\usepackage{cases}
\usepackage{cancel}
\usepackage{hyperref}
\usepackage[normalem]{ulem}
\usepackage{subfigure}
\usepackage{here}
\usepackage{color}
\usepackage{comment}

\usepackage{tikz}
\usetikzlibrary{matrix}

\allowdisplaybreaks[3]

\begin{document}
	

\title{Dynamical realization of the small field inflation \textcolor{black}{of Coleman-Weinberg type} \\
in the post supercooled universe}

\author{He-Xu Zhang}\thanks{{\tt hxzhang18@163.com}}
\affiliation{Center for Theoretical Physics and College of Physics, Jilin University, Changchun, 130012,
	China}

\author{Hiroyuki Ishida}\thanks{{\tt ishidah@pu-toyama.ac.jp}}
	\affiliation{Center for Liberal Arts and Sciences, Toyama Prefectural University, Toyama 939-0398, Japan}

\author{Shinya Matsuzaki}\thanks{{\tt synya@jlu.edu.cn}}
\affiliation{Center for Theoretical Physics and College of Physics, Jilin University, Changchun, 130012,
	China}%

\begin{abstract}

The small field inflation (SFI) of Coleman-Weinberg (CW) type suffers from precise tuning of the initial inflaton field value to be away from the true vacuum one. 
We propose a dynamical trapping mechanism to solve this problem:  
an ultra-supercooling caused by an almost scale-invariant CW potential traps the inflaton at the false vacuum, far away from the true vacuum dominantly created by the quantum scale anomaly, and allows the inflaton to dynamically start the slow-roll down due to a classical explicit-scale breaking effect. 
To be concrete, we employ a successful CW-SFI model and show that the proposed mechanism works consistently with the observed bounds on the inflation parameters. 
The proposed new mechanism thus provides new insights for developing small field inflation models. 

\end{abstract}

\maketitle


\section{Introduction}

Inflationary cosmology 
provides an elegant solution to the horizon and flatness problems, while also offering a mechanism for the generation of primordial density perturbations that seed the formation of structures. 
Among various inflation models, a class of small field inflation (SFI) based on the potential of Coleman-Weinberg (CW) type~\cite{Coleman:1973jx}, called the CW-SFI~\cite{Barenboim:2013wra},  
would be an attractive scenario because the related quantum scale anomaly 
could also be linked to the scale generation mechanism for the Standard Model possibly with the beyond the Standard Model sectors.

However, the CW-SFI possesses an intrinsic problem: in order to yield a sufficiently large e-folding number consistently with 
the observed cosmic microwave background fluctuations, the inflaton field is required to start the slow-roll (SR), away from the true vacuum, close to the top of the potential at or around the false vacuum.  
This is sort of as a fine-tuning problem, which implies necessity of a proposal for a convincing mechanism to trap the inflaton at or around the false vacuum and trigger dynamically starting the slow-roll down to the true vacuum.

The problem is simply linked to the scale invariance around the origin (the false vacuum) of the CW type potential, which is necessarily far away from the true vacuum created by the quantum scale anomaly. 
In the literature~\cite{Iso:2015wsf}, 
this intrinsic fine-tuning problem has been recapped, and 
a mechanism to trap the inflaton around the false vacuum has been proposed, in which the trapping dynamically works due to the particle number density (like plasma or a medium) created by the preheating~\cite{Dolgov:1989us,Traschen:1990sw,Kofman:1994rk,Shtanov:1994ce,Kofman:1997yn}. 
(for reviews, see, e.g., \cite{Kofman:1997yn,Amin:2014eta,Lozanov:2019jxc})~\footnote{
In a context different from the CW-SFI, the authors in~\cite{Antoniadis:2020bwi} 
have discussed another trapping idea for the initial inflaton place.  
}.

In this paper, we propose an alternative dynamical trapping mechanism. 
It is triggered by the ultra-supercooling intrinsic to the classical scale invariance
and a possible explicit scale-breaking effect on the CW-SFI, in which the latter also plays a crucial role to be fully consistent with the observational bounds on the cosmological inflation parameters, as discussed in~\cite{Iso:2014gka,Kaneta:2017lnj,Ishida:2019wkd}.

An ultra-supercooling takes place due to the delayed decay of the false vacuum, equivalently, the late tunneling interfered by the Hubble friction.  
Thus even much below the critical temperature of the CW phase transition of the first-order type, the inflaton field keeps being trapped around the false vacuum until the thermally created potential barrier gets ineffective.  
As the universe cools, the additional explicit scale-breaking term, linear in the inflaton field (with a negative slope at the origin), shifts the trapping place close to the true vacuum with holding the inflection point. 
Immediately after the inflection point goes away,
the inflaton is allowed to start the slow-roll down to the true vacuum, 
which is driven by the linear term of the explicit scale breaking.  
See Fig.~\ref{fig:veff}.

To demonstrate how the proposed trapping mechanism works practically, 
we employ a referenced CW-SFI model~\cite{Ishida:2019wkd,Miura:2018dsy} which can be thought of as a low-energy 
description of many flavor QCD with a composite dilaton as a scalon~\cite{Gildener:1976ih}, where the thermal phase transition and 
the bounce solution relevant to the supercooling are explicitly 
evaluated. 
We then show that the trapping mechanism is indeed operative consistently with the observational bounds on the cosmological inflation parameters.

\section{Scale invariant linear sigma model}

As discussed in~\cite{Miura:2018dsy}, 
the CW-type potential can be realized in a view of a linear sigma model with the classical scale symmetry along the flat direction~\cite{Gildener:1976ih}. 
This is thought of as an effective theory of an underlying large $N_f$ QCD (so-called the large $N_f$ walking gauge theory), and is compatible with the CW-SFI as shown in~\cite{Ishida:2019wkd}. 
We start with a review of the literature~\cite{Miura:2018dsy} and momentarily 
employ the linear sigma model based on the 
chiral $U(N_f)_L\times U(N_f)_R$ symmetry to derive 
the CW-type potential for the scalon~\cite{Gildener:1976ih} arising as 
the $U(N_f)$ singlet scalar meson.

The linear sigma model Lagrangian with the classical scale invariance 
takes the form 
\begin{equation}
		\mathcal{L}=\mathrm{Tr}\left[\partial_\mu M^\dagger \partial^\mu M\right]- 
  \lambda_1(\mathrm{Tr}\left[M^\dagger M\right])^2 
  - \lambda_2\mathrm{Tr}\left[(M^\dagger M)^2\right]\,.
		\label{eq:lagrangian}
\end{equation}
The linear sigma filed $M$ is decomposed into $N_f^2$ scalar mesons and $N_f^2$ pseudoscalar mesons (denoted as $s^a$ and $p^a$, respectively):
\begin{equation}
	M=\sum_{a=0}^{N_{f}^{2}-1}\left(s^{a}+i p^{a}\right) T^{a}\,,
	\label{eq:matrixfield}
\end{equation}
with $T^0=\frac{{\bf 1}}{\sqrt{2N_f}} \mathbb{I}_{_{N_f\times N_f}}$ and $T^i$ being generators of $SU(N_f)\, (i=1, \cdots,N_f^2-1)$ normalized as ${\rm Tr} [T^i T^j] = 
\delta^{ij}/2 $. 
The Lagrangian is invariant 
under $U(N_f)_L \times U(N_f)_R$ chiral transformation for $M$ as 
\begin{align}
	M \to  g_L \cdot M \cdot g_R^\dagger, \quad g_L, g_R \in U(N_f)
\,. 
\end{align} 
The $M$ is assumed to develop the vacuum expectation value (VEV) along the $U(N_f)$ singlet direction, i.e., $s^0$, which reflects the underlying 
large $N_f$ QCD nature as the vectorlike gauge theory.

Through the analysis of the renormalization group (RG) equations, 
the Gildener-Weinberg (GW) mechanism~\cite{Gildener:1976ih} tells us that if one takes the condition $\lambda_1=-\lambda_2/N_f$ at some RG scale $\mu_{\rm GW}$~\cite{Miura:2018dsy,Kikukawa:2007zk}, 
there exists a flat direction in the tree-level potential for which $V_0$ identically vanishes and a massless scalar emerges (dubbed the scalon), along which perturbation theory can be used. 
Thus, the radiative corrections along the flat direction develop a nontrivial vacuum away from the origin, 
a false vacuum as the consequence of the scale anomaly associated with the introduced RG scale. 
With a suitable renormalization condition, the one-loop potential $V_1$ 
in the present linear sigma model can thus be calculated as~\cite{Miura:2018dsy,Kikukawa:2007zk} 
\begin{widetext}
	\begin{align}
		V_1(M) = \frac{1}{64\pi^2} \sum_{a=0}^{N_f^2-1} 
		\left(
		m_{s^a}^4(M) \left( \ln{\frac{m_{s^a}^2(M)}{\mu_{_{\rm GW}}^2}} - \frac{3}{2} \right)
		+
		m_{p^a}^4(M) \left( \ln{\frac{m_{p^a}^2(M)}{\mu_{_{\rm GW}}^2}} - \frac{3}{2} \right)
		\right)+ \epsilon_0\,,
	\end{align}
\end{widetext}
where $\epsilon_0$ is a constant vacuum energy. $m_{s^a}^2$ and $m_{p^a}^2$ are the mass functions 
for scalars and pseudoscalars:  
\begin{align}
	m_{s^a}^2 = \frac{\partial^2 V_0(M)}{\partial (s^a)^2}\,, 
	\quad \quad
	m_{p^a}^2 =\frac{\partial^2 V_0(M)}{\partial (p^a)^2}\,. 
\end{align}  
By means of the chiral rotation, it is possible to choose $s^0$ to be the flat direction as 
\begin{align}
	\langle M 
	\rangle = 
	T^0 \langle s^0\rangle =\frac{1}{\sqrt{2 N_f} } \mathbb{I} \cdot  \langle s^0\rangle\,.
\end{align} 
 Then $m_{s^a}^2$ and $m_{p^a}^2$ 
 can be expressed as
\begin{align}
	\label{eq:mass}
	m_{s^0}^2(s^0) &= 0\,, 
	& 
	m_{s^i}^2(s^0) &= \left(\lambda_1 +\lambda_2\frac{3}{N_f}\right)(s^0)^2=\frac{2 \lambda_2}{N_f} (s^0)^2\,, 
	\notag \\
	m_{p^a}^2(s^0) &= 0\,, 
\end{align} 
where the flat direction condition $\lambda_1 + \lambda_2/N_f =0$ has been used. 
It is clear to see two types of the Nambu Goldstone (NG) bosons at this moment, 
where one is the scalon, $s^0$, associated with the spontaneous breaking of the scale symmetry along the flat direction, while the other corresponds to 
the NG bosons, $p^a$, for the spontaneous chiral breaking. 
Accordingly, the effective potential for the scalon $s^0$ is given by
\begin{equation}
V_{\rm eff}(s^0) = 
\frac{N_f^2-1}{64\pi^2} m_{s^i}^4(s^0)
\left( \ln{\frac{m_{s^i}^2(s^0)}{\mu_{_{\rm GW}}^2}} - \frac{3}{2} \right)+ \epsilon_0\,.
\label{eq:veff-0}
\end{equation}

We introduce an explicit chiral and scale-breaking term to the potential, 
\begin{align} 
 - c s^0 \,, 
\end{align}
which, in a sense of the underlying large $N_f$ QCD, corresponds to 
the current mass term for the hidden/dark quarks, hence makes 
the chiral NG bosons ($p^a$) pseudo. 
Then the potential of $s^0$ in Eq.(\ref{eq:veff-0}) gets shifted as  
\begin{equation}
	V_{\rm eff}(s^0)=-c\, s^0 +\frac{N_f^2-1}{64\pi^2} m_{s^i}^4(s^0)
	\left( \ln{\frac{m_{s^i}^2(s^0)}{\mu_{_{GW}}^2}} - \frac{3}{2} \right)+ \epsilon_0\,,
\label{eq:veff}
\end{equation}
where we have kept the leading order terms in perturbation series of small $c$, 
so that the $p^a$ and $s^0$ loop contributions have been dropped~\footnote{
\textcolor{black}{
We have checked that the higher order terms in powers of $c$, which arise along with the one-loop factor, do not 
affect the success of the SFI (until the inflaton $s^0$ reaches the true vacuum) within the current observation accuracy for the inflation parameters, as long as the size of $c$ (equivalently 
the size of $m_\pi$ as quoted in Eq.(\ref{para-set:1})) is small enough.} 
}. 
The stationary condition for this modified effective potential is as follows:
\begin{align}
	0&=\left.\frac{\partial V_{\rm eff}(s^0)}{\partial s^0}\right|_{s^0 \to \langle s^0 \rangle} 
	\notag \\
	&= -c +
	\frac{\lambda_2^2}{4\pi^2}
	\frac{N_f^2-1}{N_f^2}
	\langle s^0\rangle^3
	\left(
	\ln{\frac{m_{s^i}^2(\langle s^0 \rangle)}{\mu_{_{\rm GW}}^2}} - 1
	\right)\,.
\label{eq:stationary}
\end{align}
Then the VEV of $s^0$ is related to the RG scale $\mu_{GW}$ as the consequence of the dimensional transmutation: 
\begin{align}
	\mu_{_{GW}} = \sqrt{\frac{2 \lambda_2}{N_f}} \langle s^0\rangle\, e^{-\frac{2\pi^2 \cdot c \cdot N_f^2}{\lambda_2^2 (1-N_f^2)\langle s^0\rangle^3}+1}\,.
\end{align}

\section{Matching with the walking-dilaton inflaton potential}

As argued in the literature~\cite{Miura:2018dsy}, 
the scalon potential in Eq.(\ref{eq:stationary}) can be regarded 
as the composite dilaton potential arising as the nonperturbative 
scale anomaly in the underlying walking (almost scale-invariant) 
gauge theory as large $N_f$ QCD. 
In that case, the mesonic loop corrections (of ${\cal O}(1/N_c)$ in the large $N_c$ expansion) 
along the flat direction 
are matched with the nonperturbative scale anomaly term ($\sim g^2 G_{\mu\nu}^2 ={\cal O}(1/N_c)$) which, in terms of the walking  dilaton effective theory~\cite{Matsuzaki:2013eva}, takes the CW-type potential form as well.

Including the explicit chiral-scale breaking term, 
the potential of the walking dilaton inflaton $(\chi)$ takes the form~\cite{Ishida:2019wkd}
\begin{equation}
	V(\chi) = -\frac{C}{2N_f}\chi\,{\rm Tr}\left[U+U^\dagger\right]+\frac{\lambda_\chi}{4}\chi^4\left(\ln\frac{\chi}{v_\chi}+A\right)+V_0\,,
\end{equation}
with~\cite{Ishida:2019wkd}
\begin{align}
&U = e^{2\mathrm{i}\pi^i T^i/f_\pi}\,,\quad C=\frac{N_c N_f m_\pi^2 m_F^2}{8\pi^2 v_\chi}\,,\\
&\lambda_\chi\simeq \frac{16N_c N_f}{\pi^4}\left(\frac{m_F}{v_\chi}\right)^4\,, \quad A =-\frac{1}{4} +\frac{C}{\lambda_\chi v_\chi^3}\,,
\end{align}
where $V_0$ denotes a constant vacuum energy; $m_\pi$ and $f_\pi$ are the pion mass and the pion decay constant in the large $N_f$ walking gauge theory; $m_F$ is the fermion dynamical mass;  
$v_\chi$ stands for the walking dilaton inflaton VEV.

The quartic coupling $\lambda_\chi$ for the inflaton $\chi$ is required 
to be extremely tiny so as to realize the observed amplitude of the scalar perturbation. 
As was stressed in~\cite{Ishida:2019wkd},
this tiny quartic coupling can naturally be realized due to the 
intrinsic walking nature yielding a large enough scale hierarchy between $m_F$ and $v_\chi$.

Matching Eq.(\ref{eq:stationary}) with the above $V(\chi)$, 
we find the correspondence 
\begin{align}
	s^0 &= \chi\,,  \qquad \langle s^0 \rangle =\langle \chi \rangle =v_\chi \,, \qquad c =C\,, \notag \\
	\lambda_2^2 &=\frac{2\pi^2 N_f^2}{N_f^2-1}\lambda_\chi \,,
 \qquad \epsilon_0=V_0\,.
\label{correspondence}
\end{align}
Thus the free parameters in the linear sigma model can be evaluated 
in terms of the underlying large $N_f$ walking gauge theory, which makes it  possible to incorporate the thermal corrections into the walking dilaton inflation potential from the linear sigma model description, 
as noted in~\cite{Miura:2018dsy}.

The slow roll parameters ($\eta$ and $\epsilon$), the e-folding number $(N)$ 
and the magnitude of the 
scalar perturbation $(\Delta_R^2)$ are respectively defined as  
\begin{align} 
 \eta & = M_{\rm pl}^2 \left( \frac{V^{\prime \prime}(\chi)}{V(\chi)} \right) 
\,, \notag \\ 
\epsilon & = \frac{M_{\rm pl}^2}{2} \left( \frac{V^{\prime}(\chi)}{V(\chi)} \right)^2  
\,, \notag \\ 
N & = \frac{1}{M_{\rm pl}^2} \int_{\chi_{\rm end}}^{\chi_{\rm ini}} d \chi \left( \frac{V(\chi)}{V'(\chi)} \right)
\,, \notag \\ 
\Delta_R^2& = \frac{V(\chi)}{24 \pi^2 M_{\rm pl}^4 \epsilon}
\,, \label{SFI-paras}
\end{align}
with $M_{\rm pl}$ being the reduced Planck mass $\simeq 2.4 \times 10^{18}$ GeV. 
The SFI with the extremely small chiral-scale breaking by the $m_\pi$ 
will give an overall scaling for $\epsilon/\eta$ 
with the small expansion factors as 
$ 
\frac{\epsilon}{\eta} \sim \left( \frac{m_\pi}{m_F} \right)^4 \left(
 \frac{v_\chi}{\chi} \right)^2 
 $. 
Hence the inflation would be ended by reaching $\eta =1$, as long as 
$\chi/v_\chi > (m_\pi/m_F)^2$, as in the CW-SFI case. 
In the case with $m_\pi \ll \chi \ll m_F \ll v_\chi$, which is naturally realized in the present model, 
the $\eta$ and $\epsilon$ as well as 
the $\Delta_R^2$ and $N$ can further be approximated 
to be~\cite{Ishida:2019wkd}  
\begin{align} 
\eta & \simeq 24 \frac{M_{\rm pl}^2}{v_\chi^2} \frac{\chi^2}{v_\chi^2} \ln \frac{\chi^2}{v_\chi^2} 
\,, \notag \\ 
\epsilon & \simeq \frac{\pi^4}{{2}} \left( \frac{M_{\rm pl}}{v_\chi} \right)^2 
\left( \frac{m_\pi}{m_F} \right)^4 
\,, \notag \\ 
\Delta_R^2  & \simeq \frac{{2}}{\pi^{10}} 
\left( \frac{m_F}{v_\chi} \right)^4 \cdot \left( \frac{v_\chi}{M_{\rm pl}} \right)^6 
\left( \frac{m_F}{m_\pi} \right)^4 
\,, \notag \\ 
N & \simeq 
\frac{(\chi_{\rm end} - \chi_{\rm ini})}{\sqrt{2 \epsilon} M_{\rm pl}} 
\simeq 
\frac{(\chi_{\rm end} - \chi_{\rm ini}) v_\chi}{{6} \pi^2 M_{\rm pl}^2} 
\left( \frac{m_F}{m_\pi} \right)^2 
\,.  \label{approximations}
\end{align}

The conventional CW-SFI scenario sets the e-folding number $N$ 
by $\eta$, which leads to the incompatibility between $N$ and the spectral 
index $n_s= 1 + 2\eta$ in comparison with the observational values~\cite{Barenboim:2013wra,Takahashi:2013cxa}. 
As discussed in the literature~\cite{Iso:2014gka,Kaneta:2017lnj,Ishida:2019wkd}, this problem can be resolved by 
a small enough tadpole term corresponding to the $C$-term in Eq.(\ref{correspondence}), where $N$ is determined 
by $\epsilon$.

\section{Walking dilaton inflaton potential at finite temperature}

As noted above, along the flat direction, 
the thermal corrections to the walking dilaton potential 
can be evaluated by computing the thermal loops in which only 
the heavy scalar mesons $s^i$ flow. 
Taking into account also the higher loop corrections via the so-called daisy resummation, 
we thus get 
\begin{widetext}
\begin{align}
V_{\rm eff}(s^0,T) 
=&
-C\,s^0 +\frac{N_f^2-1}{64\pi^2} 
\mathcal{M}_{s^i}^4(s^0, T)
\left( \ln{\frac{
		\mathcal{M}_{s^i}^2(s^0, T)
	}{\mu_{_{GW}}^2}} - \frac{3}{2} \right) 
\notag \\
&+\frac{T^4}{2\pi^2} (N_f^2-1) J_B\left(
\mathcal{M}_{s^i}^2(s^0, T)
/T^2\right) + V_0 \,,
\label{eq:veffT}
\end{align} 
\end{widetext}
where 
$J_B(X^2) \equiv \sum_{a=0}^{N_f^2-1} 
\int_0^\infty x^2 
	\ln{\left(1-e^{-\sqrt{x^2+X^2}}\right)} dx
 $, 
and 
the $s^i$ scalar meson masses have been dressed as 
$\mathcal{M}_{s^i}^2(s^0, T) = m_{s^i}^2(s^0) + \Pi(T)$ with
\begin{align}
\Pi(T) = \frac{T^2}{6}\bigl((N_f^2 + 1)\lambda_1 + 2N_f \lambda_2\bigr)\Big|_{\lambda_1 = -\lambda_2 / N_f} \,.\label{eq:Pi}
\end{align}

\begin{figure}[t]
	\includegraphics[scale=0.95]{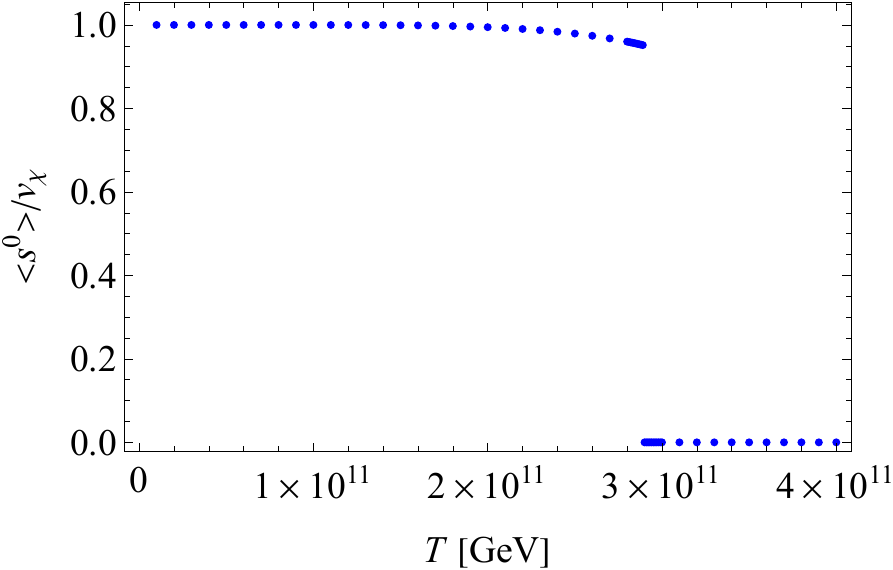}
	\caption{\label{fig:fopt} The inflaton VEV evolution with respect to temperature $T$ along the flat direction in the walking gauge theory with $N_c=3$ and $N_f=8$. }
\end{figure}

For the walking dilaton inflaton to be consistent with the observation on  cosmological inflation parameters, 
we take a benchmark parameter setting which satisfies various phenomenological and cosmological constraints~\cite{Ishida:2019wkd}  
\begin{align}
& N_c=3\,, \qquad N_f=8\,, \qquad 
v_\chi= 
1.7  
 \times 10^{15}\, {\rm GeV}\,, \notag \\ 
& m_F = 
6.2 \times 10^{11}\, {\rm GeV}\,, \qquad 
m_\pi = 
1.1 \times  10^5 \, {\rm GeV}\,, 
\label{para-set:1}
\end{align} 
which completely fixes the potential parameters in Eq.(\ref{eq:veffT}) 
through Eq.(\ref{correspondence}).  
Figure~\ref{fig:fopt} shows the $T$-dependence of the walking dilaton inflaton VEV $\langle s^0 \rangle$ which plays the role of the chiral order parameter 
in the underlying walking gauge theory. 
The extremely strong first-order phase transition is observed 
at 
\begin{align} 
T_c\simeq 3 \times 10^{11}\, {\rm GeV}
\,, \label{Tc}
\end{align} 
due to the thermally developed 
wide potential barrier between the false vacuum and the true vacuum at $\langle s^0 \rangle = v_\chi$.

\section{Trapping by supercooling and driving slow roll}

Even when the temperature cools down to $T_c$, however, the thermal chiral phase transition is not completed in the universe due to the existence of the wide barrier, hence the universe enters into a (ultra) supercooled state until the bubble nucleation takes place, in which epoch 
the potential no longer gets significant thermal corrections and almost takes the same form as the one at $T=0$.
This ultra-supercooling traps the inflaton VEV at the false vacuum until 
the inflection point of the potential goes away, that is the same timing 
as the false vacuum decay is allowed to happen. 
Accordingly, the SR starts to realize the inflation. 
In this section, we explicitly evaluate the bubble nucleation rate and 
observe this scenario in details.

The bubble nucleation rate per unit time per unit volume at high temperature is given by
\begin{equation}
	\Gamma(T) \simeq T^4 \left(\frac{S_3/T}{2\pi}\right)^{3/2}\, {\rm exp}\left(-\frac{S_3(T)}{T}\right)\,,
	\label{eq:nucleationrate}
\end{equation}
where $S_3(T)$ is the $O(3)$ symmetric bounce action determined by the following equation of motion:
\begin{align}
\label{eq:diffeq}
\frac{d^2 s_b^0(r, T)}{dr^2} + \frac{2}{r}\frac{d s_b^0(r,T)}{dr} - \frac{dV_{\text{eff}}(s_b^0,T))}{ds_b^0} = 0 \,,
\end{align}
with the boundary conditions 
\begin{align}
	\left. \frac{2}{r}\frac{d s_b^0(r)}{dr} \right|_{r=0} = 0 \,,\quad 
	s_b^0(r)|_{r=\infty} =s^0_{_{\rm FV}} \,. 
\end{align}
Here $r=0$ corresponds to the center of the bubble and $s^0_{_{\rm FV}}$ is the location of the false vacuum. The bubble nucleation temperature $T_n$ is defined at the moment when the bubble nucleation rate first catches up with the Hubble expansion rate
\begin{equation}
	\frac{\Gamma(T_n)}{H(T_n)^4}\sim 1 \,.
\end{equation}
This turns out to be amount to $S_3(T_n)/T_n \simeq 100$ via Eq.(\ref{eq:nucleationrate}) with the value of $T_n$ to be fixed later.

\textcolor{black}{Before moving on to the detailed numerical analysis, 
it is instructive to analytically understand how essentially the ultra-supercooling can be generated in the present scenario.} 
As noted by Witten in~\cite{Witten:1980ez}, near the origin and the barrier, 
the effective potential in Eq.(\ref{eq:veffT}) can be well approximated to be   
\begin{equation}
	V_{\rm eff}(s^0,T) \simeq \frac{\lambda_2(N_f^2-1)T^2}{12N_f}(s^0)^2 - \frac{N_f^2-1}{8\pi^2} 
	\frac{\lambda_2^2}{N_f^2}\ln{\frac{\mu_{_{\rm GW}}}{T}}(s^0)^4 \,,
\label{Veff:approx}
\end{equation}
where we have ignored the tadpole term, \textcolor{black}{
because it is tiny enough not to significantly affect developing the barrier and its height and width, which will be clarified later on. 
}
The tunneling rate can then analytically be evaluated as~\cite{Witten:1980ez} 
\begin{equation}
\frac{S_3(T_n)}{T_n} \simeq \frac{37.794\pi^2}{\sqrt{6}} \frac{N_f^{3/2}}{\lambda_2^{3/2}(N_f^2-1)^{1/2}}\frac{1}{\ln(\mu_{_{\rm GW}}/T_n)}\,.
\label{analytic}
\end{equation}
This implies that for small enough $\lambda_2$ as in the present almost scale-invariant model ($\lambda_2 \sim 10^{-6}$), 
the bubble nucleation temperature $T_n$ will become $\ll T_c$ and the universe experiences an ultra-supercooling~\footnote{ 
Since the tunneling definitely happens before the barrier vanishes, 
it is also reasonable to identify the nucleation temperature $T_n$ as the 
temperature $T_{\rm van}$ at which the barrier becomes vanishing, namely, 
like $T_n \simeq T_{\rm van}$. 
} 

\begin{figure}[t]
\includegraphics[scale=0.95]{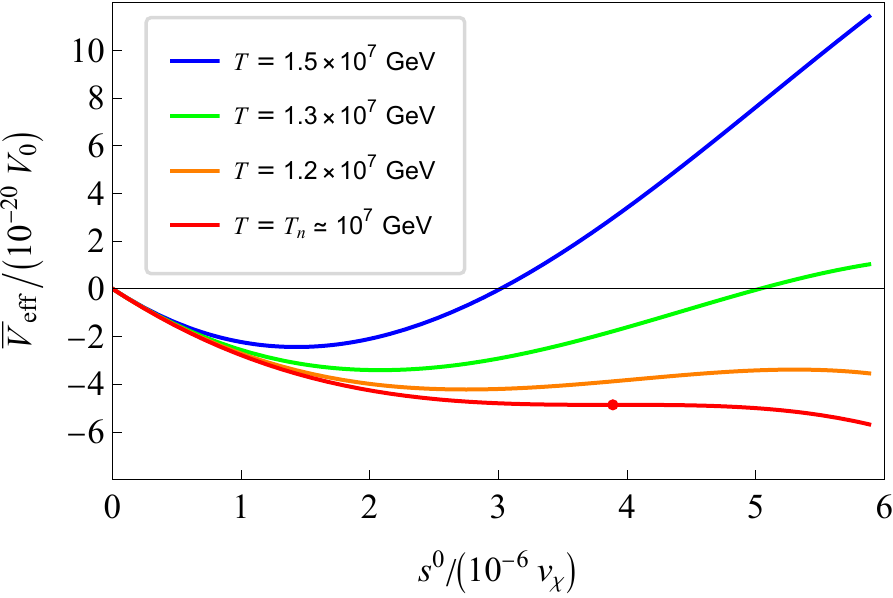}
\caption{\label{fig:veff} 
 The plot of the inflaton potential versus the inflaton field $s^0$ around the temperature $T_n =  10^7$ GeV for the false vacuum decay, until when the 
 inflaton keeps being trapped at the false vacuum $s_{\rm FV}^0$ depending on $T$. 
 The blob put on the curve at $T=T_n$ corresponds to the starting place for the inflaton to slow-roll, where the barrier and the saddle (inflection) point created by the barrier and the tadpole term at $s^0=0$ vanish. 
 The inflaton potential has been normalized as 
 $\overline{V}_{\rm eff} \equiv V_{\rm eff}(s^0, T) - V_{\rm eff}(s_0=0, T)$ with the vacuum energy $V_0$ and a scale factor of $10^{-20}$. 
 The horizontal axis has also been normalized by the inflaton VEV $v_\chi$ at the true vacuum with a scale factor of $10^{-6}$.} 
\end{figure}

The inflaton is trapped in the false vacuum created by 
the explicit-scale breaking $C$ term linear in $s^0$, in Eq.(\ref{eq:veffT}), 
and the term quadratic in $s^0$, in Eq.(\ref{Veff:approx}), which is hence  
shifted close to the true vacuum as $T$ gets lower. 
The barrier, hence the stationary false vacuum does not go away 
until the temperature reaches $T_n (\ll T_c)$. 
When the barrier and the false vacuum are gone, 
the inflaton starts to slowly roll down from the inflection point and the inflation of the universe begins. 
It is a slow enough roll, which is guaranteed by the approximate scale invariance.  
See also Fig.~\ref{fig:veff}.

Thus the starting point of the SR inflation is dynamically determined 
by the disappearance of the inflection point, when the false vacuum becomes no longer the inflection or stationary point:  
\begin{align} 
V'(s^0_{\rm FV}(T_n)) \neq 0\,, \qquad 
V''(s^0_{\rm FV}(T_n)) \neq 0 
\,. \label{slow-roll-start}
\end{align}

As the temperature cools down and gets close to $T_n$, 
the curvature at $s_{\rm FV}^0(T\searrow T_n)$ becomes smaller than (square of) the Hubble scale ($\sim \sqrt{V_0/M_{\rm pl}^2}$)~\footnote{
During the deSitter expansion era, which 
involves the supercooling epoch in the present scenario, 
the long-wave length fluctuating modes would cease the inflaton 
to settle in the (false) vacuum, when the inflaton mass gets less than the Hubble scale. 
As seen from Fig.~\ref{fig:veff}, 
however, the curvature at the $T$-dependent false vacuum $V''(s_{\rm FV}^0(T))$ keeps sizable enough compared to the 
(square of) the Hubble scale $H^2 \sim V_0/M_{\rm pl}^2 =[{\cal O}(10^2 \, {\rm GeV})]^2$. Only at the vicinity $T \sim T_n$, we have $V''(s_{\rm FV}^0 (T\sim T_n)) \lesssim H^2$. This time scale is thus too short for the long-wave length modes 
to grow (i.e., $\ln [(T_n + \Delta T)/T_n] \sim H \Delta t \ll 1$), so it cannot destabilize the false vacuum in the present scenario.  
}. 
Since the inflaton equation of motion then goes like 
$\ddot{s^0} + 3 H \dot{s^0} \approx 0$ at around $s_{\rm FV}^0(T\searrow T_n)$, 
one might think that  
it implies presence of 
an ultra slow roll (USR)~\cite{Dimopoulos:2017ged,Yi:2017mxs,Pattison:2018bct,Motohashi:2014ppa,Liu:2020oqe,Martin:2012pe,Kinney:2005vj}
before the SR starts, so that the inflation processes 
two phases: first starts with USR, and then turns to SR. 
However, it is not the case because 
$\ddot{s^0} =0 $ when the barrier disappears at $T=T_n$, 
so the motion of $s^0$ goes with $3 H \dot{s^0} = 0$, i.e., 
with zero initial velocity. Thus 
$s^0$ starts to roll with null acceleration, velocity, and curvature, 
and then, starts the SR as the curvature develops when $T$ decreases from $T_n$. 
Hence there is no extra journey for $s^0$ to experience other than the normal SR precisely at $s^0 = s_{\rm FV}^0(T_n)$, and no extra scalar perturbation generated once scalar fluctuating modes exit the Horizon when the SR inflation start there, which is identified at the pivot scale in the power spectrum. 
In fact, we have explicitly numerically 
checked that the SR condition is satisfied once $s^0$ is allowed to move from $s_{\rm FV}^0(T_n)$.


Given the parameter setting in Eq.(\ref{para-set:1}), 
\textcolor{black}{we numerically analyze the bounce solution to get }
\begin{align} 
 T_n \sim   10^7\, {\rm GeV}
\,, \qquad 
 s^0_{\rm FV}(T_n) \sim 7 \times 10^9 \, {\rm GeV} 
\,. \label{Tn}
\end{align} 
The estimated $T_n$ is indeed much lower than $T_c$ in Eq.(\ref{Tc}), due to the approximate scale invariance (with $\lambda_2 \sim 10^{-6}$), consistently with the observation based on the simplified analytic formula in Eq.(\ref{analytic}). 
The estimated $s^0_{\rm FV}$ coincides with 
the initial place of the successful walking-dilaton SR inflation~\cite{Ishida:2019wkd} with $s_{\rm FV}^0(T_n)  
\sim 7 \times 10^9$ GeV, 
where the latter was fixed merely by phenomenological constraints without taking into account the supercooling.  
We have also checked that the potential at $T=T_n$ does not substantially 
differ from the one at $T=0$ (both are still within the same order of magnitude all the way, in terms of $\overline{V}_{\rm eff}$). 
This implies that all the successful results on the SFI scenario in~\cite{Ishida:2019wkd} on the base at $T=0$ 
can simply be applied based on the conventional SFI formulae in Eq.(\ref{SFI-paras}).

The e-folding number is also accumulated during the ultra-supercooling 
when the universe cools from $T_c$ (in Eq.(\ref{Tc})) to $T_n$ (in Eq.(\ref{Tn})), in addition to $N\sim 46$ during the SR inflation epoch 
which is yielded from the formula in Eq.(\ref{SFI-paras}). 
If it is simply summed up, 
the total amount $N\sim 10+ 46 =56$ is still in good agreement with the desired e-folding to explain the universe today.

Thus the currently proposed mechanism for trapping and driving the inflaton to the SFI works, which dynamically solve the fine-tuning problem on the starting place of the SR ($s_{\rm FV}^0(T_n)/v_\chi \sim 10^{-5}$ in the present reference model), and is shown also to be consistent with the observation of the cosmological inflation parameters.

\section{Conclusion}

We have proposed a dynamical trapping mechanism to solve the problem 
on the fine-tuning of the starting place for the SR inflation, that the CW-SFI intrinsically possesses. 
The mechanism is essentially constructed from two ingredients: one is an ultra-supercooling caused by an almost scale-invariant potential of CW type, which traps the inflaton at around the false vacuum, far away from the true vacuum dominantly created by the quantum scale anomaly, while the other is 
a classical explicit-scale breaking effect, which allows the inflaton to dynamically start the SR. 
We have demonstrated how the mechanism works by employing a successful CW-SFI model and also shown the consistency with the observed bounds on the cosmological inflation parameters.

The proposed new mechanism is straightforwardly applicable to 
other models of CW type, and 
thus provides new insights for developing small field inflation models.  

 \section*{Acknowledgments} 

We are grateful to Taishi Katsuragawa for useful comments and discussion. 
This work was supported in part by the National Science Foundation of China (NSFC) under Grant No.11747308, 11975108, 12047569, 
and the Seeds Funding of Jilin University (S.M.), 
and Toyama First Bank, Ltd (H.I.).

\end{document}